\documentclass[%
twocolumn,
superscriptaddress,
showpacs,
 aps,
 prb,
10pt
]{revtex4-2}

\usepackage{mathtools}
\usepackage{amsmath}
\usepackage{amssymb}
\usepackage{bbm}
\usepackage{bm}
\usepackage{graphicx}
\usepackage{xcolor}
\usepackage{dcolumn}
\usepackage{bm}
\usepackage{float}
\usepackage{subfigure}
\usepackage[colorlinks]{hyperref}
\usepackage{dsfont}
\usepackage{braket}
\usepackage{units}
\usepackage{placeins}

\newcommand{\mat}[1]{\bm{#1}}
\newcommand{\matK}[1]{\underline{#1}} 
\renewcommand{\vec}[1]{\bm{#1}}

\DeclareMathOperator*{\IIm}{Im}
\newcommand{\ee}{\mathrm{e}}  
\newcommand{\ii}{\mathrm{i}} 
\DeclareMathOperator*{\Tr}{Tr} 
\DeclareMathOperator*{\dd}{d} 
\begin{document}
\title{Auxiliary master equation approach to the Anderson-Holstein impurity problem out of equilibrium}
\author{Daniel Werner}
\email[]{daniel.werner96@posteo.at}
\affiliation{Institute of Theoretical and Computational Physics, Graz University of Technology, 8010 Graz, Austria}
\author{Rok Žitko}
\email[]{rok.zitko@ijs.si}
\affiliation{Jo\v{z}ef Stefan Institute, Jamova 39, SI-1000 Ljubljana, Slovenia}
\affiliation{Faculty of Mathematics and Physics, University of Ljubljana, Jadranska 19, SI-1000 Ljubljana, Slovenia}
\author{Enrico Arrigoni}
\email[]{arrigoni@tugraz.at}
\affiliation{Institute of Theoretical and Computational Physics, Graz University of Technology, 8010 Graz, Austria}

\begin{abstract}
We introduce a method based on auxiliary master equation for solving the problem of an impurity with local electron-electron and electron-phonon interaction embedded between two conduction leads with a finite bias voltage. The Anderson-Holstein Hamiltonian is transformed to a corresponding Lindblad equation with a reduced set of sites providing an optimal approximation of the hybridization function. The problem is solved in the superfermion representation using configuration interaction for fermions and bosons. The phonon basis is shifted and rotated with the intention of permitting a low phonon basis cutoff even in the strong-coupling regime. We benchmark this approach with the numerical renormalization group in equilibrium, finding excellent agreement. We observe, however, that the rotation brings no advantage beyond the bare shift. This is even more apparent out of equilibrium, where issues in convergence with respect to the size of the phononic Hilbert space occur only in the rotated basis. As an application of the method,  we explore the evolution of the phononic peak in the differential conductance spectra with changing phonon frequency. 
\end{abstract}

\date{\today}
\maketitle

\section{Introduction}

Molecular electronics is the endeavor of using single molecules as components in ultra-miniaturized circuits, making use of their particular electronic and vibration properties \cite{jaklevic1966,lambe1968,aviram1974,reed1997,chen1999onoff,nitzan2001,donhauser2001,zhitenev2002,qiu2003,yu2004c,galperin2007,zimbovskaya2011,song2011,aradhya2013}. Non-equilibrium properties of single molecules can be probed by embedding molecules in gaps between electrodes \cite{park2000}, in mechanically controlled break junctions \cite{agrait2003}, as well as using scanning tunneling spectroscopy \cite{stipe1998,moresco2004}. In these setups it is possible to apply bias voltages that are large compared to characteristic energy scales of the problem, such as inter-level spacing, phonon frequency, and electron-electron repulsion. Spectroscopic measurements are, however, difficult to interpret because the strongly-interacting problems are difficult to solve reliably and accurately out of equilibrium. A number of theoretical tools has been developed in the past~\cite{paaske2005,pi.gr.23.ss, galperin2007prb, we.mi.07, PhysRevB.101.125405, rourabas2013,PhysRevB.93.115139,laakso2014,PhysRevB.101.235439}, but no method has emerged yet as the ultimate solution applicable to all parameter domains. 

In this work we present the auxiliary master equation approach for the Anderson-Holstein impurity problem \cite{holstein1959,anderson1961,newns1969,jeon2003,hewson2002,cornaglia2004}. This Hamiltonian is the minimal description of a molecule with a single low-energy orbital with effective on-site electron-electron repulsion and with the coupling between the on-site charge and the displacement of a local vibration mode, embedded between two metallic conduction leads. The goal is to solve this problem in the regime of sizable electron-electron (e-e) and electron-phonon (e-ph) coupling for large bias voltages and to calculate the differential conductance, which is the main experimentally measurable quantity that provides information on the excitations in the system.

The paper is structured as follows. In Sec.~\ref{sec:methods} we start by discussing the basic idea of the auxiliary master equation approach (AMEA). We continue by briefly reviewing our implementation of configuration interaction (CI) for the electrons in Sec.~\ref{sec:CI_electron}, and discuss in more detail the CI treatment of phonons in Sec.~\ref{sec:CI_phonon}. In Sec.~\ref{sec:results} we first elaborate on the choice of method parameters. Then we benchmark the conductance against the numerical renormalisation group in equilibrium in Sec.~\ref{sec:eq_results}, and show conductance results out of equilibrium in Sec.~\ref{sec:noneq_res}.

\section{\label{sec:methods}Model and Method}
\subsection{\label{sec:non_eq_gf}Non-equilibrium Green's functions}
We use the Keldysh formalism~\cite{schw.61, kad.baym, keld.65, ra.sm.86, ha.ja, wagn.91}. The time contour from $t \rightarrow -\infty$ to $t \rightarrow \infty$ and back again leads to a $2\times2$ matrix structure for including all time combinations, i.e. both times on the upper contour, one above and one below, etc. Since we are only interested in the steady state, we can take advantage of time translation invariance and fix one time argument of the Green's functions to zero. This allows us to work directly in the frequency domain instead of time domain. The general form of a Green's function is then 
\begin{equation}\label{eqn:GF_full}
    \matK{G}(\omega) = 
    \left (\begin{matrix}
    G^\text{R}(\omega) & G^\text{K}(\omega) \\
    0 & G^\text{A}(\omega)
    \end{matrix}
    \right ),
\end{equation}
where $G^A = \left ( G^\text{R} \right ) ^\dagger$. Throughout this paper, underlined quantities represent a $2\times2$ structure in Keldysh space. In equilibrium, $G^\text{K}$ can be determined from $G^\text{R}$ via the fluctuation-dissipation theorem as
\begin{equation}\label{eqn:fluc_diss}
    G^\text{K}(\omega) = F(\omega)2i\IIm[G^\text{R}(\omega)],
\end{equation}
with
\begin{equation}\label{eqn:b_f_function}
    F(\omega) = \begin{cases} 
\coth{\frac{\beta \omega}{2}} & \text{for bosons,} \\
\tanh{\frac{\beta \omega}{2}} & \text{for fermions.} \\
\end{cases}
\end{equation}
\subsection{\label{sec:imp_phys}Physical impurity model}
\begin{figure}
\centering
\includegraphics[scale=0.7]{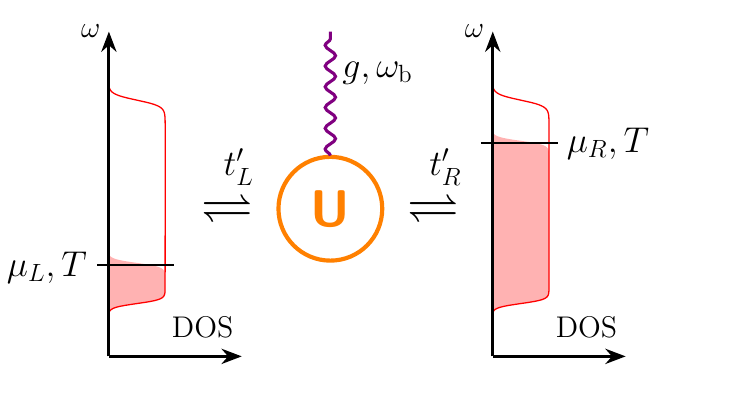}
\caption{Schematic representation of the system. Two leads at different chemical potentials are connected to an impurity with a Hubbard interaction $U$ and a coupling to a Holstein phonon.}
\label{fig:schematic}
\end{figure}

The physical setup consists of an impurity, two electronic reservoirs and one local phonon mode, as depicted in Fig.~\ref{fig:schematic}. The respective Hamiltonian may be split up as
\begin{equation}
    H = H_\text{imp} + H_\text{bath} + H_\text{coup}.
\end{equation}
$H_\text{imp}$ is the Hamiltonian of the impurity,
\begin{equation}
    \varepsilon_\text{imp} \sum_\sigma d_\sigma^\dagger d_\sigma + U n_{d \uparrow} n_{d \downarrow} + \sum_\sigma g d_{\sigma}^{\dagger} d_{\sigma} (b + b^\dagger) + \omega_\text{b} b^\dagger b,
\end{equation}
with Hubbard interaction $U$, on-site energy $\varepsilon_\text{imp}$, creation (annihilation) operator $d^\dag_\sigma$ ($d_\sigma$) of a fermion at the impurity site with spin $\sigma$, particle number operator $n_{d \sigma}$, electron-phonon interaction strength $g$, and phonon frequency $\omega_\text{b}$. The leads are described by 
\begin{equation}
    H_\text{bath} = \sum_{k \lambda \sigma} \varepsilon_{\lambda k} a^\dagger_{\lambda k \sigma} a_{\lambda k \sigma}
\end{equation}
with dispersion $\varepsilon_{\lambda k}$ and creation (annihilation) operator $a_{\lambda k \sigma}^\dagger$ ($a_{\lambda k \sigma}$) of a fermion in the left and right lead, $\lambda \in \{L, R\}$, labeled by momentum $k$. The coupling between the impurity and the bath is given by
\begin{equation}
    H_\text{coup} = \frac{1}{\sqrt{N_k}}\sum_{k \lambda \sigma} t_\lambda' (a^\dagger_{\lambda k \sigma} d_\sigma + d_\sigma^\dagger a_{\lambda k \sigma}),
\end{equation}
where $t'_\lambda$ is the coupling strength between the leads and the impurity, and $N_k \rightarrow \infty$ is the number of $k$ points. Whenever not mentioned otherwise we have $t_L' = t_R'$.

Alternatively, the environment may be be described in terms of Green's functions by defining the hybridization function ~\cite{economou} 
\begin{equation} \label{eqn:hyb_phy}
    \matK{\Delta}_\mathrm{ph} = \sum_\lambda t_\lambda'^{2} \matK{g}_\lambda(\omega),
\end{equation}
where $\matK{g}_\lambda(\omega)$ are the Green's functions of the decoupled leads. (The subscript $\mathrm{ph}$ here stands for ``physical'', to distinguish this hybridization function from the auxiliary function to be defined in the next section.) To fully define the problem, one needs to specify the band properties. We chose a flat density of states smoothed around the edges so that
\begin{equation} \label{eqn:lead_gfs}
    -\IIm[g^R_\lambda] = \frac{\pi}{2 D} \rho_\text{FD}(\omega - D, T_\text{fict}) \rho_\text{FD}(-\omega - D, T_\text{fict}),
\end{equation}
where $\rho_\text{FD}$ is the Fermi-Dirac distribution, $T_\text{fict}$ a fictitious temperature, and $D$ the half band-width. The smoothing is introduced to avoid a sharp change that would reduce the quality of reproducing the environment within AMEA; $T_\text{fict}$ merely quantifies the degree of smoothing. The parameters are set to $T_\text{fict} = 0.5 \Gamma$ and $D = 10 \Gamma$ throughout this paper. Whenever not mentioned otherwise, the coupling between the reservoir and the impurity is set to $t_\lambda' = \sqrt{\Gamma D / \pi} = \Gamma \sqrt{10/\pi}$ (which gives $-\IIm[\Delta_\text{ph}(\omega = 0)] = \Gamma$).

Since the uncoupled leads themselves are in equilibrium, the Keldysh part of their Green's function can be calculated using the fluctuation-dissipation theorem, Eqs.~\eqref{eqn:fluc_diss},\eqref{eqn:b_f_function}. If not stated otherwise, the chemical potentials of the right and left reservoir are given as $\mu_\text{R} = -\mu_\text{L} = \phi/2$, i.e., $\phi=eV$ describes the voltage drop across the impurity.
\subsection{\label{sec:imp_aux}Auxiliary impurity model}
Since the physical model is defined in an infinitely large Hilbert space, one has to find a way of approximating it for numerical computations on a reduced set of sites. The basic idea of the AMEA is to set the parameters of the Lindblad equation so as to reproduce the physical hybridization function~\cite{do.so.17}. 

The Lindblad equation is given as
\begin{equation}\label{eqn:Lindlad}
    \begin{split}
        \frac{\dd \rho(t)} {\dd t} &= \hat{\hat{L}}\rho       \\
        &= -i[H_\text{aux}, \rho] + \sum_{ij\sigma} \Gamma_{ij}^{(1)} \left ( c_{j\sigma} \rho c_{i\sigma}^\dagger - \frac{1}{2} \{c_{i\sigma}^\dagger c_{j\sigma}, \rho \} \right ) 
        \\
        &+ \sum_{ij} \Gamma_{ij}^{(2)} \left ( c_{i\sigma}^\dagger \rho c_{j\sigma} - \frac{1}{2} \{c_{j\sigma} c_{i\sigma}^\dagger, \rho \} \right ),
    \end{split}
\end{equation}
where $c_{i\sigma}^\dagger$ $(c_{i\sigma})$ is the fermionic creation (annihilation) operator, $\rho$ is the density matrix, $\Gamma_{ij}^{(1)}$ and $\Gamma_{ij}^{(2)}$ describe the dissipative contributions, and $H_\text{aux}$ is the unitary part, given as
\begin{equation} \label{eqn:Haux}
    \begin{split}
        H_\text{aux} &= \sum_{\substack{ \sigma \\ i, j: \langle ij \rangle \wedge (i,j) \neq (0,0)}} E_{ij} c_{i\sigma}^\dagger c_{j\sigma} + U n_{f\uparrow}n_{f\downarrow} \\
        &+ \sum_{\sigma} g c_{f \sigma}^{\dagger} c_{f \sigma} (b + b^\dagger) + \omega_\text{b} b^\dagger b \\
        &= \underbrace{\sum_{\substack{\sigma \\ i, j: \langle ij \rangle \wedge i, j \neq 0}} E_{ij} c_{i\sigma}^\dagger c_{j\sigma}}_{{\textstyle H_{\text{aux env}}\mathstrut}} + \underbrace{\sum_{i \in \{-1, 1\}, \sigma} E_{fi} c_{i\sigma}^\dagger c_{f\sigma} + h.c.}_{{\textstyle H_{\text{aux coup}}\mathstrut}} \\
        &\begin{drcases}
        & + \sum_\sigma E_{ff} c_{f\sigma}^\dagger c_{f\sigma} + g c_{f \sigma}^{\dagger} c_{f \sigma} (b + b^\dagger)
        \\
        &+ U n_{f\uparrow}n_{f\downarrow} + \omega_\text{b} b^\dagger b.
        \end{drcases} H_{\text{aux imp}}
    \end{split}
\end{equation}
We define $f:=0$, and the indices $i$ and $j$ can assume integer values from $-N_\text{B}/2$ to $N_\text{B}/2$, where $N_\text{B}$ is the number of bath sites. In this paper we only consider cases in which $N_\text{B} = 6$. This number of bath sites already allows for a large number of parameters, which can be seen when considering that the $\mat{\Gamma}$ matrices are only restricted to be positive semidefinite. In other words, the number of parameters used to fit $\Delta_\text{aux}$ to $\Delta_\text{ph}$ increases  quadratically in $N_B$. In Ref.~\onlinecite{do.so.17} it is furthermore explicitly shown that $N_B = 6$ suffices to reproduced a flat density of states accurately.

The part of $H_\text{aux}$ denoted as $H_{\text{aux imp}}$ is equivalent to $H_\text{imp}$. $H_{\text{aux coup}}$ describes the coupling between the auxiliary reservoirs and the impurity, i.e. its parameters are set by $t'_\lambda$. All the other contributions appearing in Eq.~\eqref{eqn:Haux} are determined in a fitting procedure, which we will discuss in the following.

The auxiliary hybridization function $\Delta_\mathrm{aux}$ is defined through
\begin{equation} \label{eqn:hyb_aux}
    \begin{split}
        \matK{G}_{\text{0}ff} &= \left (\matK{g}_\text{0} - \matK{\Delta}_{\text{aux}} \right )^{-1} \rightarrow
        \\
        \Delta_{\text{aux}}^\text{R}(\omega) &= 1/g_{\text{0},ff}^\text{R}(\omega) - 1/G_{\text{0},ff}^\text{R}(\omega),
        \\
        \Delta_{\text{aux}}^\text{K}(\omega) &= G_{\text{0},ff}^\text{K}(\omega)/|{G_{\text{0},ff}^\text{R}(\omega)}|^2,
    \end{split}
\end{equation}
where $\matK{G}_{\text{0}ff}$ is given~\cite{do.so.17} by
\begin{equation} \label{eqn:G_0_aux}
    \begin{split}
        \mat{G}_\text{0}^\text{R}(\omega) &= [\omega - \mat{E} + i (\mat{\Gamma}^{\text{(1)}} + \mat{\Gamma}^{\text{(2)}})]^{-1},
        \\
        \mat{G}_\text{0}^\text{K}(\omega) &= 2 i \mat{G}_\text{0}^\text{R}(\omega)(\mat{\Gamma}^{\text{(2)}} - \mat{\Gamma}^{\text{(1)}})\mat{G}_\text{0}^\text{A}(\omega).
    \end{split}
\end{equation}
and $g_{0, ff}^R$ as
\begin{equation} \label{eqn:g_0_aux}
    g_{0, ff}^R(\omega) = (\omega - \varepsilon_\text{imp})^{-1}.
\end{equation}
A note on notation: Bold quantities indicate the $(N_\text{B}+1) \times (N_\text{B}+1)$ structure in the space of auxiliary levels, lower case $g$ indicates the decoupled setup, upper case $G$ contains the coupling, and the index $0$ indicates the non-interacting case.

Given the physical hybridization function $\Delta_\mathrm{ph}$, one defines a cost function
\begin{equation}
    \begin{split}
        \chi(\mat{E},\mat{\Gamma}^{\text{(1)}},\mat{\Gamma}^{\text{(2)}}) &= \sum_{{\alpha\in\{\text{R},\text{K}\}}}\quad\int_{{-\infty}}^{{\infty}}\dd\omega\, W^{\alpha}(\omega)\times \notag \\
        &\hspace{-1.0cm}\times \IIm[\Delta_{\text{ph}}^{\alpha}(\omega) - \Delta_{\text{aux}}^{\alpha}(\omega;\mat{E},\mat{\Gamma}^{\text{(1)}},\mat{\Gamma}^{\text{(2)}})]^{2}
    \end{split}
\end{equation}
which must be minimized to obtain the parameters leading to the optimal approximation of the physical hybridization function.

In this paper we set the weight function to $W^\alpha(\omega) = \Theta(|{\omega} - \omega_\text{max}|)$, with $\omega_\text{max}/\Gamma = 15$. This range suffices to capture the hybridization function appropriately, since our flat density of states ranges from $\omega/\Gamma = -10$ to $10$, and decays exponentially outside this region. A more detailed discussion about the fitting procedure can be found in Ref.~\onlinecite{do.so.17}.
\subsection{\label{sec:superferm}Superfermion representation}
\begin{figure}
\centering
\includegraphics[scale=0.4]{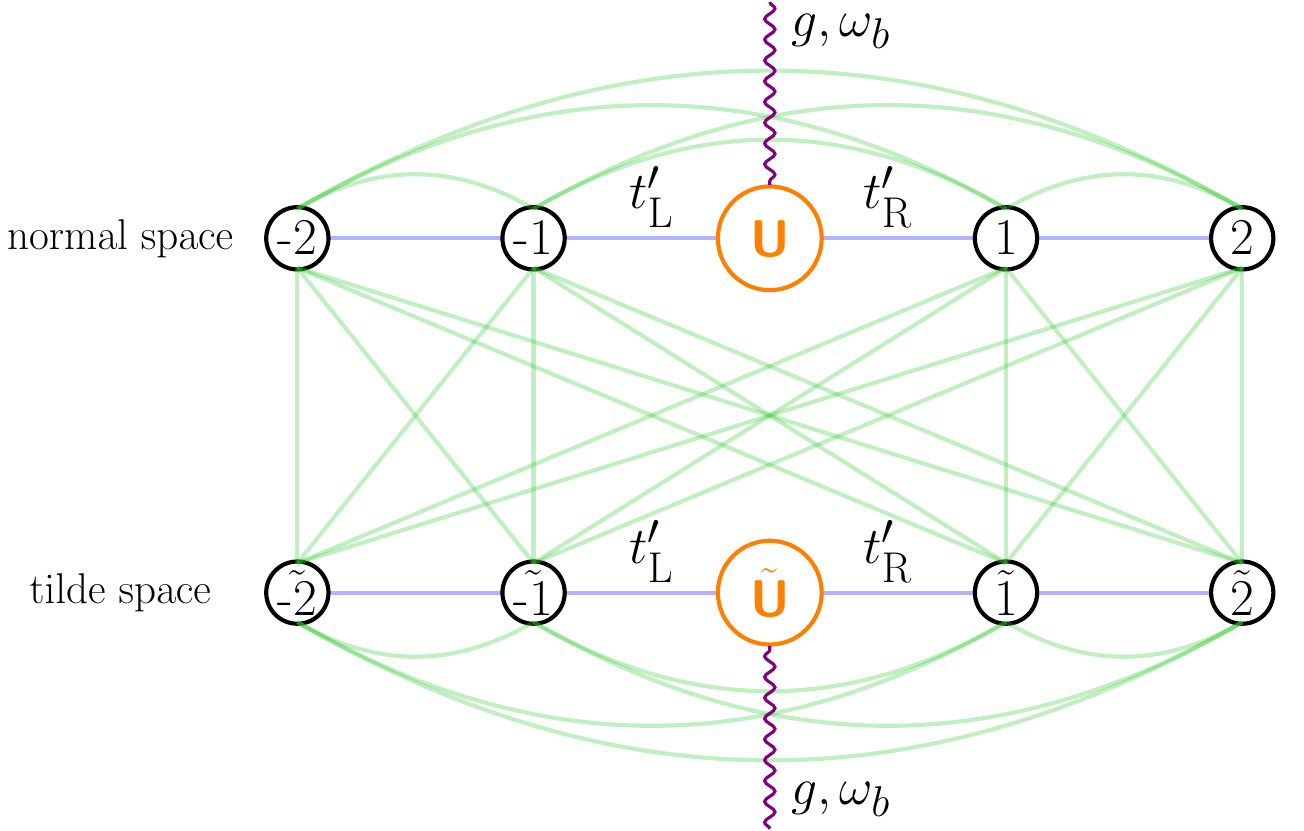}
\caption{Schematic representation of the auxiliary system.}
\label{fig:aux}
\end{figure}
For computational reasons, one may transform the Lindblad equation into the superfermion representation~\cite{dz.ko.11}.  In this form the Lindbladian becomes a matrix and the density matrix a vector. In this section we will briefly sketch the basic ideas of this procedure, roughly following Ref.~\cite{dz.ko.11}, see also~\cite{do.nu.14, ar.do.18}.

We introduce the left vacuum~\footnote{The left vacuum is always a left eigenstate with eigenvalue zero for arbitrary Lindbladians in superfermion representation. Furthermore it is important when calculating observables, since it corresponds to taking the trace of the density matrix it is multiplied with.} defined as
\begin{equation} \label{eqn:left_vac}
    \ket{I} = \sum_{\{m_\text{el}, m_\text{ph}\}} \underbrace{(\ket{m_\text{el}} \otimes \ket{m_\text{ph}})}_{\text{normal space}} \otimes \underbrace{(\ket{\Tilde{m_\text{el}}} \otimes \ket{\Tilde{m_\text{ph}}})}_{\text{tilde space}},
\end{equation}
where $m_\text{el}$ and $m_\text{ph}$ run over all states in the respective Hilbert spaces. Here we doubled the Hilbert space by introducing the "tilde" space. 

Then we apply the density matrix on the left vacuum
\begin{equation} \label{eqn:density_matrix_superfermion}
    \begin{split}
        \rho \to \ket{\rho} &= (\rho \otimes \tilde{\mathds{1}}) \ket{I} \\
        &= \left ( \sum_{mn} \rho_{mn} \ket{m}\bra{n} \otimes \tilde{\mathds{1}} \right ) \left( \sum_{j} \ket{j} \otimes \ket{\tilde{j}} \right) \\
        &= \sum_{mnj} \rho_{mn} (\ket{m}\braket{n|j}) \otimes \ket{\tilde{j}} = \sum_{mn} \rho_{mn} \ket{m}\otimes\ket{\tilde{n}}.
    \end{split}
\end{equation}

This yields a vector in the doubled Hilbert space that contains all information of the density matrix. 
As the next step, the Lindbladian must also be transformed accordingly.
The aim is to replace all density matrices appearing in $L$ with $\ket{\rho}$. Therefore, we calculate $L \ket{I}$ and whenever $\ket{I}$ is next to $\rho$ we can use Eq.~\eqref{eqn:density_matrix_superfermion}. 
To achieve this, one uses the tilde-conjugation rules, given as
\begin{equation} \label{eqn:tilde_conj_rel}
    \begin{split}
        c_j^\dagger \ket{I} &= - \ii \tilde{c}_j^{} \ket{I}, \\
        c_j \ket{I}^{} &= - \ii \tilde{c}_j^\dagger \ket{I}, \\
        b^\dagger \ket{I} &= \tilde{b} \ket{I}, \\
        b \ket{I}^{} &= \tilde{b}^\dagger \ket{I},
    \end{split}
\end{equation}
since the creation and annihilation operators, once transferred into the tilde space, commute with the density matrix~\footnote{The density matrix has an even number of normal-space operators.} and anticommute (fermionic) or commute (bosonic) with the normal-space operators. The Lindbladian then takes the form
\begin{equation} \label{eqn:Lindblad_superferm}
    \begin{split}
        \ii L &= \sum_\sigma \vec{c}_\sigma^\dagger \mat{h} \vec{c}_\sigma^{} -2 \Tr(\mat{E}+\ii\mat{\Lambda}) \\
        &+U \left ( n_{f\uparrow}^{} n_{f\downarrow}^{} - \tilde{n}_{f\uparrow}^{} \tilde{n}_{f\downarrow}^{} \right ) \\ &\underbrace{+\omega_\text{b} (b^\dag b - \Tilde{b}^\dag \Tilde{b}) + g (c_{f \sigma}^\dag c_{f \sigma} (b + b^\dag) - \Tilde{c}_{f \sigma}^\dag \Tilde{c}_{f \sigma} (\Tilde{b} + \Tilde{b}^\dag))}_{\ii L_\text{eph}},
    \end{split}
\end{equation}
with the matrix
\begin{equation} \label{eqn:h_matrix}
    \begin{split}
        \mat{h} = \left(\begin{matrix}
        \mat{E} + \ii \mat{\Omega} & \phantom{-}2 \mat{\Gamma}^{\text{(2)}} \\
        -2 \mat{\Gamma}^{\text{(1)}} & \mat{E} - \ii \mat{\Omega}
        \end{matrix}\right),
    \end{split}
\end{equation}
where $\vec{c}_\sigma^\dagger = (c_{-N_{\text{B}}/2, \sigma}^\dagger,\ldots , c_{N_{\text{B}}/2, \sigma}^\dagger, \tilde{c}_{-N_{\text{B}}/2, \sigma}^{},\ldots , \tilde{c}_{N_{\text{B}}/2, \sigma}^{})$, $\mat{\Omega} = \mat{\Gamma}^{\text{(2)}} - \mat{\Gamma}^{\text{(1)}}$ and $\mat{\Lambda} = \mat{\Gamma}^{\text{(2)}} + \mat{\Gamma}^{\text{(1)}}$. As defined in Sec.~\ref{sec:imp_aux}, the impurity site is indexed  $f := 0$. Fig.~\ref{fig:aux} shows the Lindblad setup in its superfermion form~\footnote{Ref.~\onlinecite{do.so.17} shows that under usage of all available parameters, as is the case here, the geometry has no influence on the result. This justifies our choice here.}, where the blue lines represent the unitary contributions ($\mat{E}$) and the green ones the dissipative contributions ($\mat{\Gamma}^{(1/2)}$). Furthermore, since the Lindbladian in Eq. \eqref{eqn:Lindblad_superferm} is not a superoperator anymore, it does not carry the double hat. This also allows to easily distinguish between superoperators and operators in superfermion space.

As can be deduced from Eq.~\eqref{eqn:Lindblad_superferm}, normal and tilde fermionic particles are always created and annihilated simultaneously, therefore their difference is conserved, i.e.,
\begin{equation} \label{eqn:conservation}
    N_{\sigma} - \tilde{N}_{\sigma} = \sum_{i} (c_{i\sigma}^{\dagger}c_{i\sigma}-\tilde{c}_{i\sigma}^{\dagger}\tilde{c}_{i\sigma}).
\end{equation}
This is valid for both spins separately, since angular momentum is also conserved.

Since the left vacuum is always a left eigenstate of the Lindbladian with eigenvalue zero, it lies in the subspace $N_{\sigma}-\tilde{N}_{\sigma} = 0$. This can be observed from Eq.~\eqref{eqn:left_vac}, since all fermionic states have by definition the same number of normal and tilde electrons. Since the corresponding right eigenvector is the steady state, it is restricted to the same subspace.

The fermionic Green's functions can be expressed in the Lehmann representation. For positive times ($+$) the greater ($>$) and lesser ($<$) component are given as
\begin{equation} \label{eqn:Lehmann_greater}
    \begin{split}
        G_{ij}^{>+}(\omega) &= \sum_k \bra{I} c_i^{} \ket{k\text{R}} \bra{k\text{L}} c_j^\dagger \ket{\rho_\infty} \frac{1}{\omega - \ii L_k^{}}, \\
        G_{ij}^{<+}(\omega) &= \sum_k \bra{I} c_j^{\dagger} \ket{k\text{R}} \bra{k\text{L}} c_i^{} \ket{\rho_\infty} \frac{1}{\omega + \ii L_k^{}},
    \end{split}
\end{equation}
where $\bra{k\text{L}}$ and $\ket{k\text{R}}$ are the left and right eigenvectors of the Lindbladian and $L_k$ the corresponding eigenvalues. From this all other fermionic Green's functions of interest, as well as the self energy ($\Sigma$), can be obtained:
\begin{equation} \label{eqn:GF_relations}
    \begin{split}
        G_{ij}^{\gtrless-}(\omega) &= -[G_{ji}^{\gtrless+}(\omega)]^{\ast}, \\
        G_{ij}^{\text{R}}(\omega) &= G_{ij}^{>+} - G_{ij}^{<-}, \\
        G_{ij}^{\text{K}}(\omega) &= G_{ij}^{>+} + G_{ij}^{<+} - G_{ij}^{>-} - G_{ij}^{<-}, \\
        \Sigma^{\text{R}}(\omega) &= 1/G_\text{0}^\text{R}(\omega) - 1/G_{}^\text{R}(\omega), \\
        \Sigma^{\text{K}}(\omega) &= -G_\text{0}^K(\omega)/|G_\text{0}^\text{R}(\omega)|^2 + G_{}^\text{K}(\omega)/|G_{}^\text{R}(\omega)|^2.
    \end{split}
\end{equation}
For bosonic Green's functions there exist similar expressions as Eqs. \eqref{eqn:Lehmann_greater} and \eqref{eqn:GF_relations}. Computationally, we calculate the steady state using the biconjugate gradient method, and the Green's function using the Bi-Lanczos scheme~\cite{lanczos}. The basis used to express $L$ is obtained as described in the next section.
\subsection{\label{sec:CI}Configuration Interaction}
The most straightforward basis choice for $L$ is in terms of $c/c^\dagger$ and $b/b^\dagger$ operators. Then one can perform the calculation using the full many-body Lindbladian using a cutoff in the number of included phonons. The resulting matrix grows exponentially in the system size $N_B$, therefore we rotate the bosonic and fermionic single particle operators with the aim of cutting off the fermionic states in a controlled way, and obtaining a reduced cutoff in the bosonic states. This is discussed in the following subsections, where we roughly follow Ref.~\onlinecite{dz.ko.14}.
\subsubsection{\label{sec:CI_electron}Electrons}
The treatment of electrons in the AMEA using CI is explained in detail in Ref.~\onlinecite{we.lo.23}. Here we will only give a brief overview of the main steps. We start by defining
\begin{equation}
    \begin{split}
        E_{ij\sigma} &= E_{ij} + U\braket{n_{f\Bar{\sigma}}}\delta_{if}\delta_{jf} - 2 \frac{g^2 \braket{n_f}}{\omega_\text{b}}, \\
        \mat{h}_\sigma &= \left(\begin{matrix}
        \mat{E}_\sigma + \ii \mat{\Omega} & \phantom{-}2 \mat{\Gamma}^{\text{(2)}} \\
        -2 \mat{\Gamma}^{\text{(1)}} & \mat{E}_\sigma - \ii \mat{\Omega}
        \end{matrix}\right),
    \end{split}
\end{equation}
where $\braket{n_{f \sigma}} = \bra{I} c_{f \sigma}^\dagger c_{f \sigma} \ket{\rho_\infty}$ and $\braket{n_f} = \braket{n_{f \uparrow}} + \braket{n_{f \downarrow}}$. Here $\mat{E}_\sigma$ (and thereby $\mat{h}_\sigma$) takes into account the electron-electron and electron-phonon (the origin of the phononic term will become clear in the following, see Sec. \ref{sec:CI_phonon}) interaction  on a mean field level.

The matrix $\mat{E}_\sigma$ is in principle the same for both spins, since there is no magnetic term in the Lindbladian. We showed in Ref.~\onlinecite{we.lo.23} that choosing a spin-dependent (magnetised) Hartree-Fock term reduces the error originating from the basis cutoff we introduce within CI. The respective mean field (Hartree-Fock) Lindbladian reads
\begin{equation} \label{eqn:Lindblad_superferm_nonint}
    \ii L_\text{0el} = \sum_\sigma \vec{c}_\sigma^\dagger \mat{h}_\sigma \vec{c}_\sigma^{} \underbrace{ - \sum_\sigma \Tr(\mat{E}_\sigma+\ii\mat{\Lambda})}_\eta.
\end{equation}
In our previous paper we fixed $\braket{n_{f\uparrow}} = 0.3$ and $\braket{n_{f\downarrow}} = 0.7$ to introduce an artificial magnetisation. The exact values for $\braket{n_{f \sigma}}$ do not have a strong influence on the shape of the impurity Green's function, as long as it is far enough from $\braket{n_{f \uparrow}} = \braket{n_{f \downarrow}}$, but not too far (roughly in the range 0.6-0.9). Since not all the results we show here are particle-hole symmetric, we compute the total Hartree-Fock electron occupation self-consistently using Eq.~\eqref{eqn:Lindblad_superferm_nonint} and introduce a magnetisation as $\braket{n_{f \uparrow}} = 0.3 \braket{n_f}$, $\braket{n_{f \downarrow}} = 0.7 \braket{n_f}$.

The non-Hermitian matrix $\mat{h}_\sigma$ can be diagonalised straightforwardly as
\begin{equation}
    \mat{\varepsilon}_\sigma = \mat{V}_\sigma^{-1} \mat{h}_\sigma \mat{V}_\sigma,
\end{equation}
where $\mat{V}_\sigma$ ($\mat{V}^{-1}_\sigma$) are the right (left) eigenvectors of $\mat{h}_\sigma$, and $\mat{\varepsilon}_\sigma$ its eigenvalues. In this basis, the Hartree-Fock Lindbladian reads
\begin{equation}
    \ii L_\text{0el} = \sum_\sigma \Bar{\mat{\xi}}_\sigma \mat{\varepsilon}_\sigma \mat{\xi}_\sigma + \eta.
\end{equation}
The new operators are defined as $\Bar{\mat{\xi}}_\sigma = \mat{c}_\sigma^\dag \mat{V}_\sigma$ and $\mat{\xi}_\sigma = \mat{V}_\sigma^{-1} \mat{c}_\sigma$, which obey fermionic anticommutation relations. However, the creation and annihilation operators are not hermitian conjugates of each other, i.e. $(\xi)^\dagger \neq \Bar{\xi}$. 

The steady state of the non-interacting Lindbladian $L_\text{0el}$ can be found by identifying which operators annihilate it. In other words, we must have 
\begin{equation}
    \begin{split}
        \xi_{i \sigma} \ket{\rho_\text{$\infty$0el}} = 0\quad\text{ for }\IIm(\varepsilon_{i \sigma})<0, \\
        \Bar{\xi}_{i \sigma} \ket{\rho_\text{$\infty$0el}} = 0\quad\text{ for }\IIm(\varepsilon_{i \sigma})>0,
    \end{split}
\end{equation}
because anything else would imply a divergent state, as becomes apparent when considering
\begin{equation} \label{eqn:time_evolution}
    \ee^{L_0 t} \xi_{i \sigma} \ket{\rho_\text{$\infty$0el}} = \ee^{L_0 t} \xi_{i \sigma} \ee^{-L_0 t} \ket{\rho_\text{$\infty$0el}} = \ee^{\ii\varepsilon_{i \sigma} t} \xi_{i \sigma} \ket{\rho_\text{$\infty$0el}}.
\end{equation}
For computational reasons we furthermore perform a particle-hole transformation

\begin{equation} \label{eqn:op_transformation}
\begin{split}
\vec{P} &= \Bar{\mat{D}} \vec{\xi} + \Bar{\vec{\xi}} \mat{D}, \\
\Bar{\vec{P}} &= \mat{D} \vec{\xi} + \Bar{\vec{\xi}} \Bar{\mat{D}}
\end{split}
\end{equation}
with components
\begin{equation} \label{eqn:D_imag}
\begin{split}
D_{ij}^{} &= \delta_{ij}^{} \Theta[\IIm(\varepsilon_i^{})], \\
\Bar{D}_{ij}^{} &= 1-D_{ij}^{}.
\end{split}
\end{equation}

From this we get $\ket{\rho_\text{0el}} = \ket{0}$. Starting from the Hartree-Fock steady state as a reference state, we can create a subspace of excited states by applying operator pairs, 
\begin{equation} \label{eqn:CI_apply_op}
     \Bar{\xi}_{i \sigma} \xi_{j \sigma} \ket{\rho_\text{$\infty$0el}},
\end{equation}
which is equivalent to
\begin{equation}
    \Bar{P}_{i \sigma} \Bar{P}_{j \sigma} \ket{0},
\end{equation}
with $i$ and $j$ taking all values resulting in non-vanishing states, which obey the conservation rules, namely $N_\sigma - \Tilde{N}_\sigma = c_\sigma$, where $c_\sigma$ is a spin-dependent constant. For the steady state we have $c_\downarrow = c_\uparrow = 0$ and for the Green's functions $c_{\downarrow/\uparrow} = 0$ and $c_{\uparrow/\downarrow} = \pm 1$.

\begin{figure}[t] 
\centering
\includegraphics[scale=0.5]{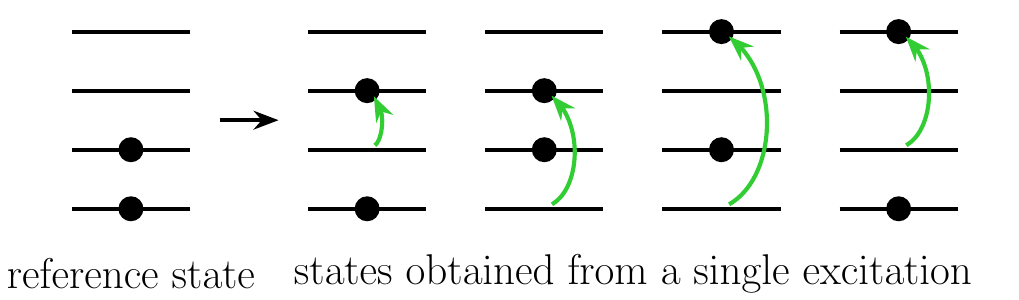}
\caption{Illustration of Eq.~\eqref{eqn:CI_apply_op} for a small Hilbert space. The reference state, in our case the steady state of Eq.~\eqref{eqn:Lindblad_superferm_nonint}, is shown on the left hand site. On its right hand site all the states one obtains from a single excitation are depicted.}
\label{fig:CI_illustration}
\end{figure}
The state in Eq.~\eqref{eqn:CI_apply_op} corresponds to a single "particle-hole" excitation, and can be graphically interpreted as depicted in Fig.~\ref{fig:CI_illustration} (shown in the $\xi$ basis). By applying more such pairs in sequence, one obtains higher excitations. In this paper we always use a basis created by up to three particle-hole excitations (referred to in the literature as CISDT). 

Eventually Eq.~\eqref{eqn:Lindblad_superferm} is transformed into the $P$ basis, such that the fermionic contributions to the matrix elements in the many-body Lindblad matrix can be calculated in this subspace.

\subsubsection{\label{sec:CI_phonon}Phonons}

When including phonons, the Hilbert space becomes in principle infinite, so that one has to introduce a cutoff in the maximum phonon number. The larger the cutoff, the better the approximation for observables, but at the cost of higher memory usage and longer computation times. By choosing an advantageous single particle basis, the number of bosonic particles can be kept low while still obtaining accurate results.

One starts by introducing an offset to the bosonic operators, which corresponds to the NECI* shift introduced in Ref.~\cite{dz.ko.14}, whose treatment we roughly follow below. This shifts the number of phonons in the new vacuum to the amount corresponding to the mean-field electron occupation, i.e., without taking into account the effect of electronic fluctuations:
\begin{equation}\label{eqn:transform_offset}
    \begin{split}
        b &\rightarrow b - \frac{g \braket{n_{f}}}{\omega_\text{b}}, \quad b^\dagger \rightarrow b^\dagger - \frac{g \braket{n_{f}}}{\omega_\text{b}}, \\
        \Tilde{b} &\rightarrow \Tilde{b} - \frac{g \braket{n_{f}}}{\omega_\text{b}}, \quad \Tilde{b}^\dagger \rightarrow \Tilde{b}^\dagger - \frac{g \braket{n_{f}}}{\omega_\text{b}}.
    \end{split}
\end{equation}
The corresponding Lindblad term in the superfermion representation then reads
\begin{equation}\label{eqn:L_phe_superfermion_offset_tr}
    \begin{split}
        i L_\text{eph} &= \sum_\sigma \frac{2 g^2 \braket{n_{f}}}{\omega_\text{b}} (\Tilde{c}_{f \sigma}^{\dagger} \Tilde{c}_{f \sigma} - c_{f \sigma}^{\dagger} c_{f \sigma}) + \omega_\text{b} (b^\dagger b - \Tilde{b}^\dagger \Tilde{b}) \\
        &+ g (b + b^\dagger)  (c_{f \sigma}^{\dagger} c_{f \sigma} - \braket{n_{f}}) \\
        &+ g (\Tilde{b} + \Tilde{b}^\dagger) (\braket{n_{f}} - \Tilde{c}_{f \sigma}^{\dagger} \Tilde{c}_{f \sigma}).
    \end{split}
\end{equation}

In the next step one "rotates" the phononic basis, i.e., one introduces a linear transformation between the "tilde" and "non-tilde" phononic creation and annihilation operator, as first introduced in Ref.~\onlinecite{dz.ko.14}.
This transformation is obtained by the exact solution of an auxiliary Lindblad equation acting on the isolated phonon level
\begin{equation}\label{eqn:ph_Lindblad}
    \begin{split}
        \hat{\hat{L}}_\text{0ph} \rho &= \alpha_+ \left(b^\dagger \rho b - \frac{1}{2} \{\rho, b b^\dagger\}\right) + \alpha_- \left(b \rho b^\dagger - \frac{1}{2} \{\rho, b^\dagger b\}\right)
        \\
        &+ \ii [b b^\dagger \omega_b, \rho],
    \end{split}
\end{equation}

this equation can be obtained exactly by eliminating the ``fermionic reservoir'' from the isolated phonon level within the weak-coupling ($g/\omega_b\ll 1$) regime with the Born-Markov-Secular approximation~\cite{scha,br.pe}. In this case,  the ratio $\frac{\alpha_-}{\alpha_+}$ can be written in terms a suitable fermionic density density correlation function, and in the equilibrium case can be inferred from the relation

\begin{equation}\label{eqn:phonon_exp_val}
     \Tr \{b^\dagger b \rho_\text{$\infty$0ph} \} = \frac{\alpha_+}{\alpha_- - \alpha_+} \rightarrow \exp(\beta \omega_\text{b}) = \frac{\alpha_-}{\alpha_+}.
\end{equation}

With this we rewrite the impurity Lindbladian $L_\text{imp}$ as
\begin{equation}
    \begin{split}
        L_\text{imp} &= L_\text{int} + L_\text{onsite} + L_\text{HF}
        \\
        L_\text{HF} &= L_\text{0el} + L_\text{0ph}
        \\
        \ii L_\text{int} &= U n_\uparrow n_\downarrow + \ii L_\text{eph} - \ii L_\text{HF},
        \\
        \ii L_\text{onsite} &= \varepsilon_\text{imp} + \sum_\sigma \frac{2 g^2 \braket{n_{f}}}{\omega_\text{b}} (\Tilde{c}_{f \sigma}^{\dagger} \Tilde{c}_{f \sigma} - c_{f \sigma}^{\dagger} c_{f \sigma}),
    \end{split}
\end{equation}
with $L_\text{0ph}$ being the superfermion version of $\hat{\hat{L}}_\text{0ph}$ given as
\begin{equation}\label{eqn:L_wc_superfermion_introduce}
    \begin{split}
        L_\text{0ph} &= \alpha_+ \left (b^\dagger \Tilde{b} - \frac{1}{2} \Tilde{b} \Tilde{b}^\dagger - \frac{1}{2} b^\dagger b \right) \\
        &+ \alpha_- \left (\Tilde{b} b - \frac{1}{2} \Tilde{b} \Tilde{b}^\dagger - \frac{1}{2} b^\dagger b \right) - i \omega_\text{b} (b^\dagger b - \Tilde{b} \Tilde{b}^\dagger) \\
        &+ \frac{\alpha_- - \alpha_+}{2} - i \omega_\text{b}.
    \end{split}
\end{equation}

$L_\text{0el}$ contains the mean-field on-site energies felt by the electrons due to the e-e and e-ph coupling and $L_\text{0ph}$ considers the thermalization effect the electrons have on the phonons. $L_\text{int}$ then contains corrections beyond the mean-field level felt by the electrons, as well as corrections to the thermalization effect the electrons have on the phonons.

In the case of the electrons, we used the matrices diagonalising $L_\text{0el}$ to transform the electronic single particle operators. For the phonons we proceed in a similar way, where we start by rewriting $L_\text{0ph}$
\begin{equation}\label{eqn:L_wc_superfermion_transform}
    \begin{split}
        L_\text{0ph} &= \mat{b}^\dagger \mat{h}_\text{0ph} \mat{b} + \eta_\text{0ph} \\
        &= \underbrace{\mat{b}^\dagger \mat{S} \mat{U} \mat{S}}_{\overline{\bm{\varphi}}} \underbrace{\mat{S} \mat{U}^{-1} \mat{S} \mat{h}_\text{0ph} \mat{U}}_{\bm{\varepsilon}} \underbrace{\mat{U}^{-1} \mat{b}}_{\bm{\varphi}} + \eta_\text{0ph},
    \end{split}
\end{equation}
where
\begin{equation}\label{eqn:L_wc_superfermion_definitions}
    \begin{split}
        &\mat{b} = \left ( \begin{matrix} b^\dagger & \Tilde{b} \end{matrix} \right ), \\
        &\eta_\text{0ph} = \frac{\alpha_- - \alpha_+}{2} - i \omega_\text{b}, \\
        &\mat{h}_\text{0ph} = \left ( \begin{matrix} -\frac{1}{2}(\alpha_+ + \alpha_-) - i \omega_\text{b} & \alpha_+ \\ \alpha_- & -\frac{1}{2}(\alpha_+ + \alpha_-) + i \omega_\text{b} \end{matrix} \right ), \\
        &\mat{S} = [\mat{b}, \mat{b^\dagger}] = \left ( \begin{matrix}
            1 & 0 \\
            0 & -1
        \end{matrix} \right ),
    \end{split}
\end{equation}
and $\mat{U}$ are the right eigenvectors of $\mat{S} \mat{h}_\text{0ph}$. For convenience we define
\begin{equation}\label{eqn:redefine_ph_op}
        \psi_1 := \phi_1, \Bar{\psi}_1 := \Bar{\phi}_1, \psi_2 := \Bar{\phi}_2, \Bar{\psi_2} := \phi_2
\end{equation}
such that
\begin{equation}
    [\mat{\psi}, \mat{\Bar{\psi}}] = \mathds{1}.
\end{equation}
In the $\psi$-basis the Lindbladian $L_\text{0ph}$ is diagonal, therefore the steady state is given by a single Fock state, which is simultaneously the vacuum in this basis ($\ket{\rho_\text{$\infty$0ph}} = \ket{00}$)~\footnote{Also the left vacuum becomes the vacuum, i.e. $\bra{I} = \bra{00}$, which it remains for arbitrary Lindbladians in this basis.}. This state contains the thermalizing effect the electrons have on the phonons, making it a good choice for a reference state~\footnote{Which is of importance if one restricts the Hilbert space, otherwise any arbitrary reference state would be equally good.} to build the phononic basis from:
\begin{equation}
    (\Bar{\psi}_1)^n (\Bar{\psi}_2)^m \ket{0}.
\end{equation}
We define the cutoff value for the number of included phonons per site as "$N_\text{ph,max}$" (Number of phonons), i.e., $0 \le n, m \le \text{$N_\text{ph,max}$}$, with $n, m$ integers. To transform Eq.~\eqref{eqn:L_phe_superfermion_offset_tr} in the $\psi$ basis we use
\begin{equation}\label{eqn:b_full_tr}
    \begin{split}
        \bm{b} &= \left ( \begin{matrix} \psi_1 + \frac{\alpha_+}{\alpha_-} \Bar{\psi}_2 \\ \psi_1 + \Bar{\psi}_2 \end{matrix} \right ) \\ 
        \bm{b^\dagger} &= \left ( \begin{matrix} \Bar{\psi}_1 + \psi_2 & \frac{\alpha_+}{\alpha_-} \Bar{\psi}_1 + \psi_2 \end{matrix} \right ) \frac{\alpha_-}{\alpha_- - \alpha_+},
    \end{split}
\end{equation}
which follows from Eqs.~\eqref{eqn:L_wc_superfermion_transform} and \eqref{eqn:redefine_ph_op}. With this $L_\text{eph}$ in its final version reads
\begin{equation}\label{eqn:L_phe_superfermion_full_tr}
    \begin{split}
        &\ii L_{\text{eph}} = \sum_\sigma 2\frac{g^2 \braket{n_{f}}}{\omega_\text{b}} (c_{f \sigma}^{\dagger} c_{f \sigma} - \Tilde{c}_{f \sigma}^{\dagger} \Tilde{c}_{f \sigma}) \\
        &+ g (c_{f \sigma}^{\dagger} c_{f \sigma} - \braket{n_{f}}) \\
        &\left ( \frac{\alpha_-}{\alpha_- - \alpha_+} \left( \Bar{\Psi}_1 + \Psi_2 \right ) + \Psi_1 + \frac{\alpha_+}{\alpha_-} \Bar{\Psi_2} \right ) \\
        &+ g (\braket{n_{f}} - \Tilde{c}_{f \sigma}^{\dagger} \Tilde{c}_{f \sigma})\\
        &\left (\Psi_1 + \Bar{\Psi}_2 + \frac{1}{\alpha_- - \alpha_+} \left( \alpha_+ \Bar{\Psi}_1 + \alpha_- \Psi_2 \right ) \right ) \\
        &+ \omega_\text{b} (\overline{\Psi}_2 \Psi_2 - \overline{\Psi}_1 \Psi_1).
    \end{split}
\end{equation}

\section{\label{sec:results}Results}

\begin{figure}
    \centering
    \includegraphics[width=0.4\textwidth]{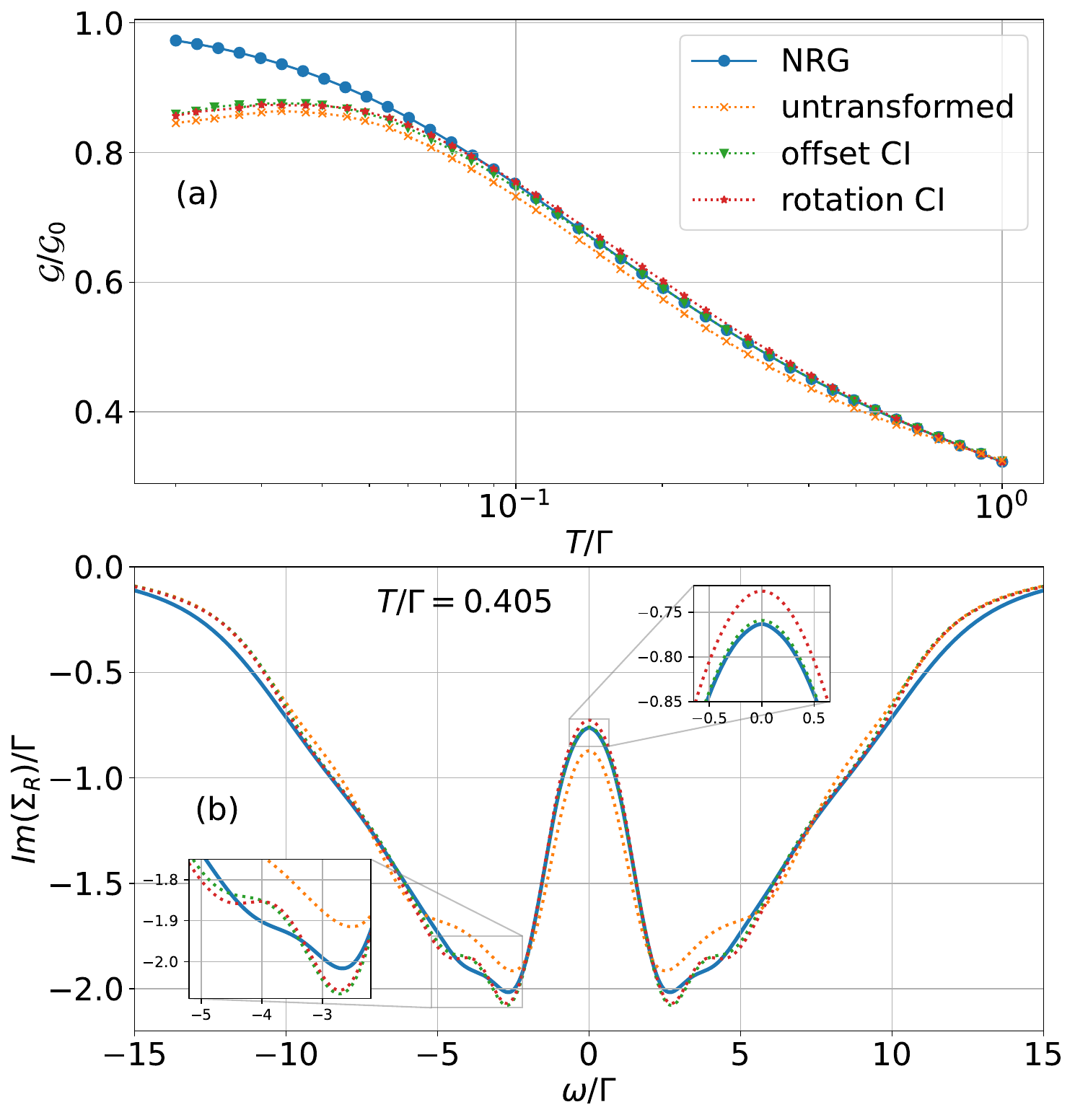}
    \caption{(a) Zero bias conductance and (b) self energy ($\IIm(\Sigma_\text{R})$) obtained with different transformations explained in the text. Parameters are $U/\Gamma = 6$, $g/\Gamma = 1.3$, $\omega_\text{b}/\Gamma = 1.3$, $\varepsilon_\text{imp} = -U/2 + 2 g^2/\omega_\text{b}$ and $N_\text{ph,max} = 5$.}
    \label{fig:eq_transf_eff}
\end{figure}

\begin{figure}
    \centering
    \includegraphics[width=.9\linewidth]{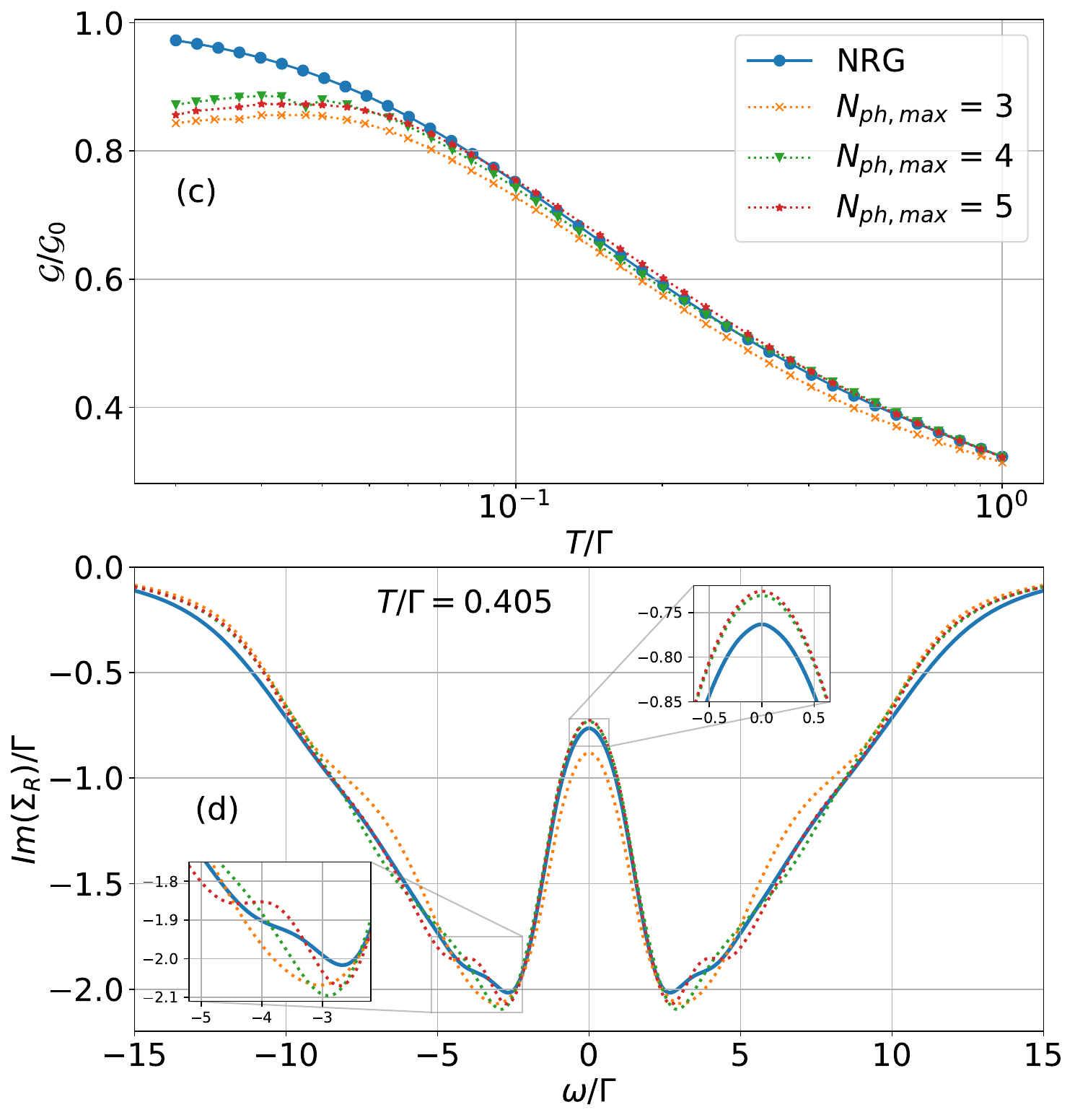}
    \hfill
    \includegraphics[width=.9\linewidth]{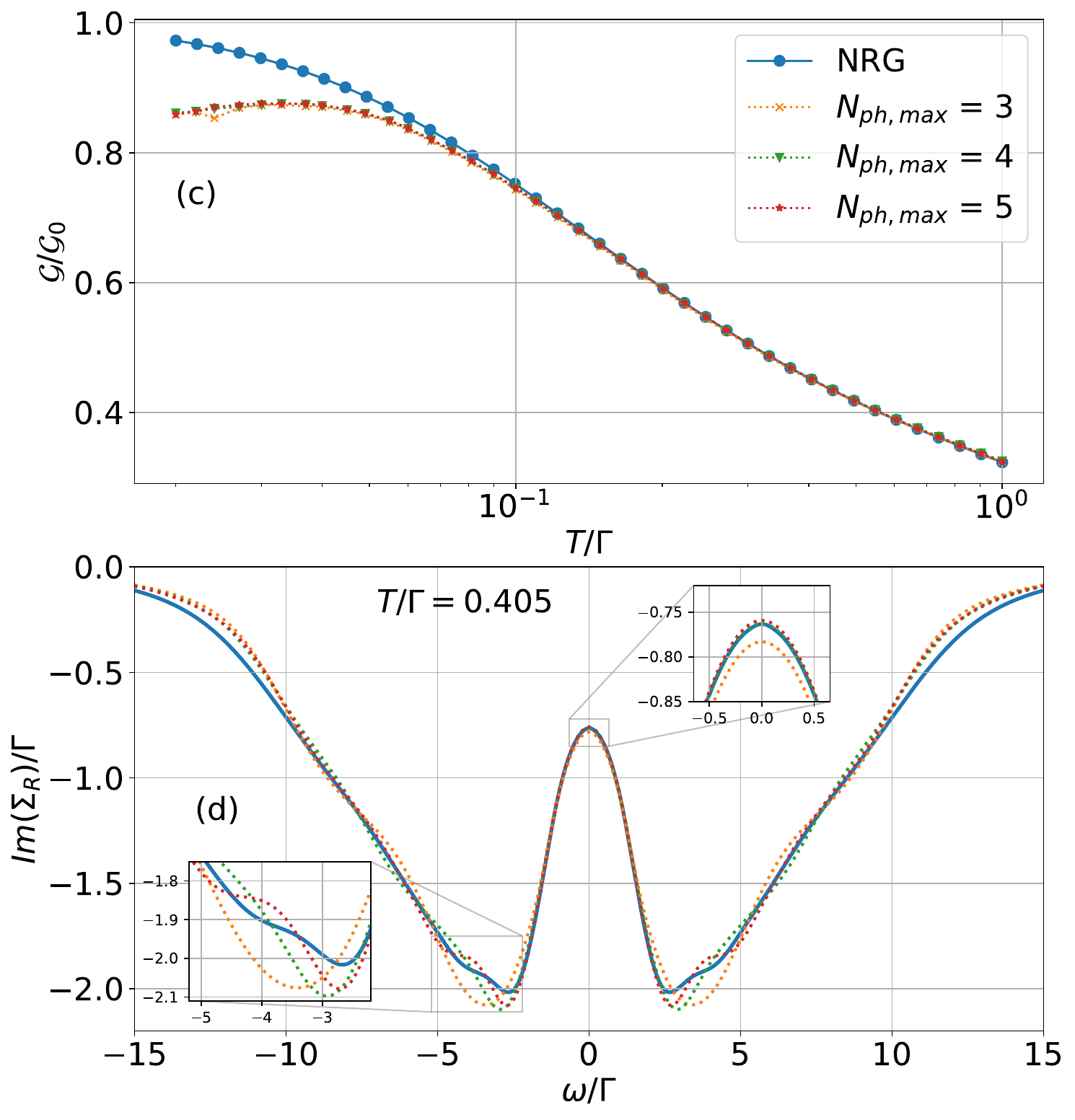}
    \caption{
    Zero bias conductance (a) and (c) and selfenergy (b) and (d) for the CI calculation with different values of the phonon cutoff compared with NRG. Results are obtained with the rotation CI (a) and (b), offset CI (c) and (d). Parameters are as in Fig.~\ref{fig:eq_transf_eff}}.
    \label{fig:eq_sps_conv}
\end{figure}

\subsection{\label{sec:model_para}Method parameters}

In equilibrium, the parameters $\alpha_+$ and $\alpha_-$ can be calculated straightforwardly from the temperature of the system. Only the ratio $\frac{\alpha_-}{\alpha_+}$ appears in equations, thus there is actually a single free parameter that is fixed by $\beta$. Out of equilibrium, we need to choose an effective temperature felt by the phonons, which is, however, not defined unambiguously. Two approaches were considered: a) fitting the Fermi function ($\rho_\text{FD}(\omega - \mu, T)$) to the non-equilibrium distribution ($\rho_\text{non-eq}(\omega)$), b) enforcing $\rho_\text{FD}(w = \omega_\text{b}) = \rho_\text{non-eq}(w = \omega_\text{b})$. Comparing the results obtained from both procedures showed that fitting works better.
Here we calculate the non-equilibrium distribution $\rho_\text{non-eq}(\omega)$ from Eq.~\eqref{eqn:Lindblad_superferm_nonint}, i.e., for the mean-field case. This is computationally much less costly than performing an iterative many-body calculation to obtain this parameter.

\subsection{\label{sec:eq_results}Comparison with the NRG in equilibrium}

As a benchmark of the approach presented here, we compare our results with NRG in the equilibrium case, where the latter is known to be very accurate, especially in the low-energy regions. First, we investigate how the transformation of the phononic basis introduced in Sec.~ \ref{sec:CI_phonon} affects the accuracy of the calculation. More specifically, we compare the results obtained without any transformation, with only the offset ("offset CI"), Eq.~\eqref{eqn:transform_offset}, and with both, the offset and the rotation ("rotation CI"), Eqs.~\eqref{eqn:b_full_tr} and \eqref{eqn:transform_offset}. We plot the zero-bias conductance as a function of temperature, which gives us insight into the effect of the transformations at low energies over a wide temperature range. While the conductance mainly probes the low-energy region, we address higher energies by evaluating $\IIm[\Sigma_R(\omega)]$. From the Meir-Wingreen formula~\cite{me.wi.92} one obtains the conductance $\mathcal{G}$ as
\begin{equation}\label{eqn:conductance}
    \begin{split}
        \mathcal{G} &= \int_{-\infty}^\infty \frac{d \omega}{\pi} \IIm[G^\text{R}(\omega, \phi)] \\
        & \times \frac{\gamma_L(\omega) \gamma_R(\omega)}{\gamma_L(\omega) + \gamma_R(\omega)} \left ( \frac{d\rho_{F, L}(\omega, \mu_\text{L}, T)}{d \phi} - \frac{d\rho_{F, R}(\omega, \mu_\text{R}, T)}{d \phi} \right ) \\
        +& \int_{-\infty}^\infty \frac{d \omega}{\pi} \frac{d \IIm[G^\text{R}(\omega, \phi)]}{d \phi} \\
        & \times \frac{\gamma_L(\omega) \gamma_R(\omega)}{\gamma_L(\omega) + \gamma_R(\omega)} \left (\rho_{F, L}(\omega, \mu_\text{L}, T) - \rho_{F, R}(\omega, \mu_\text{R}, T)\right ) \\
    \end{split}
\end{equation}
for the case of proportional coupling. Here, $\gamma_\lambda (\omega) = -2|t'^2_\lambda| \IIm[g_\lambda^R(\omega)]$ are also referred to as the "lead self-energies". We have $\gamma_R (\omega = 0) = \gamma_L (\omega = 0) = \Gamma$ unless otherwise specified (this follows from our definition of $\matK{g}_\lambda(\omega)$ in eq. \eqref{eqn:lead_gfs} and our default values for $t'_\lambda$, as given in Sec. \ref{sec:imp_phys}).

The Numerical Renormalization Group (NRG)~\cite{wils.75, bu.co.08, zi.pr.09} is well known for capturing the equilibrium properties very accurately. Therefore, we use the results obtained from NRG Ljubljana~\footnote{http://nrgljubljana.ijs.si/} as a reference for the equilibrium case \footnote{The NRG calculations have been performed using the discretization parameter $\Lambda=2$, with twist averaging over four discretization grids, keeping up to 10000 spin multiplets in truncation, and using broadening parameter $\alpha=0.2$. The final spectral functions have been computed using the self-energy trick.}. Both the offset CI and the rotation CI results compare very well with NRG and perform significantly better than the results obtained without any transformation, as can be seen in the conductance results in Fig.~\ref{fig:eq_transf_eff}(a). Considering a maximum allowed deviation of 3\% with respect to the NRG, temperatures down to 0.05$\Gamma$ can be reached for $U/\Gamma = 6$ with both CI transformations. The offset CI performs marginally better. 

The self energies of the offset and rotation CI are also almost on top of each other, see Fig.~\ref{fig:eq_transf_eff}(b). They mostly coincide with the NRG self energy, except close to the phononic feature, which is more pronounced for the CI methods. We have compared the self energies at a temperature which is close to the one obtained for non-equilibrium at $\phi = \omega_\text{b}$, using the procedure discussed in Sec.~\ref{sec:model_para}.

With increasing phonon cutoff, we expect convergence of all observables. We investigate the cutoff dependence by again plotting the zero-bias conductance as a function of temperature as well as the self energy, see Fig.~\ref{fig:eq_sps_conv}. The plots shows that the convergence behavior of the offset CI at low energies is superior. In the self energy, however, close to the phononic feature both approaches have trouble reaching full convergence.

\subsection{\label{sec:noneq_res}Non-equilibrium results}

\begin{figure}
    \centering
    \includegraphics[width=.9\linewidth]{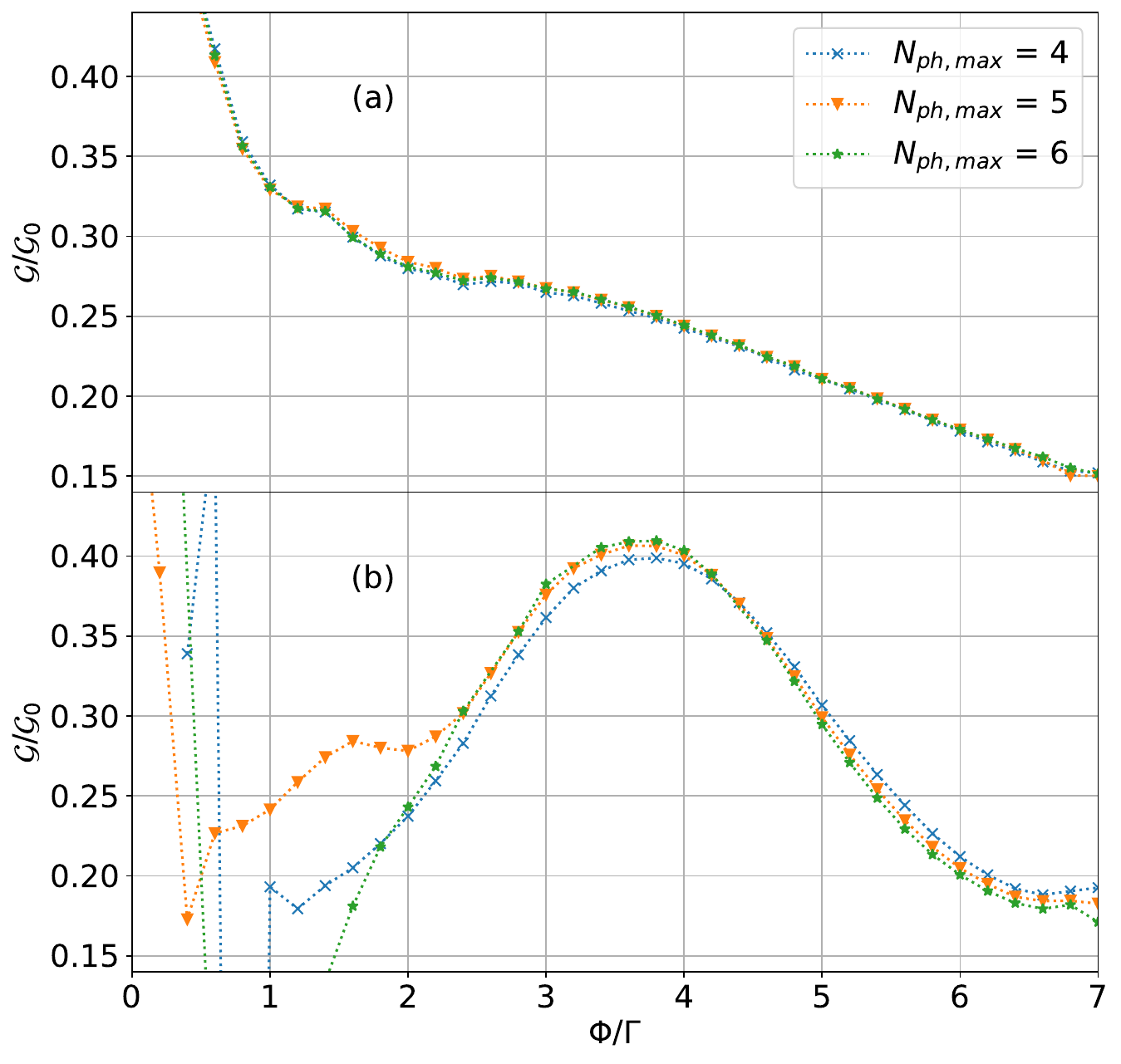}
    \hfill
    \includegraphics[width=.9\linewidth]{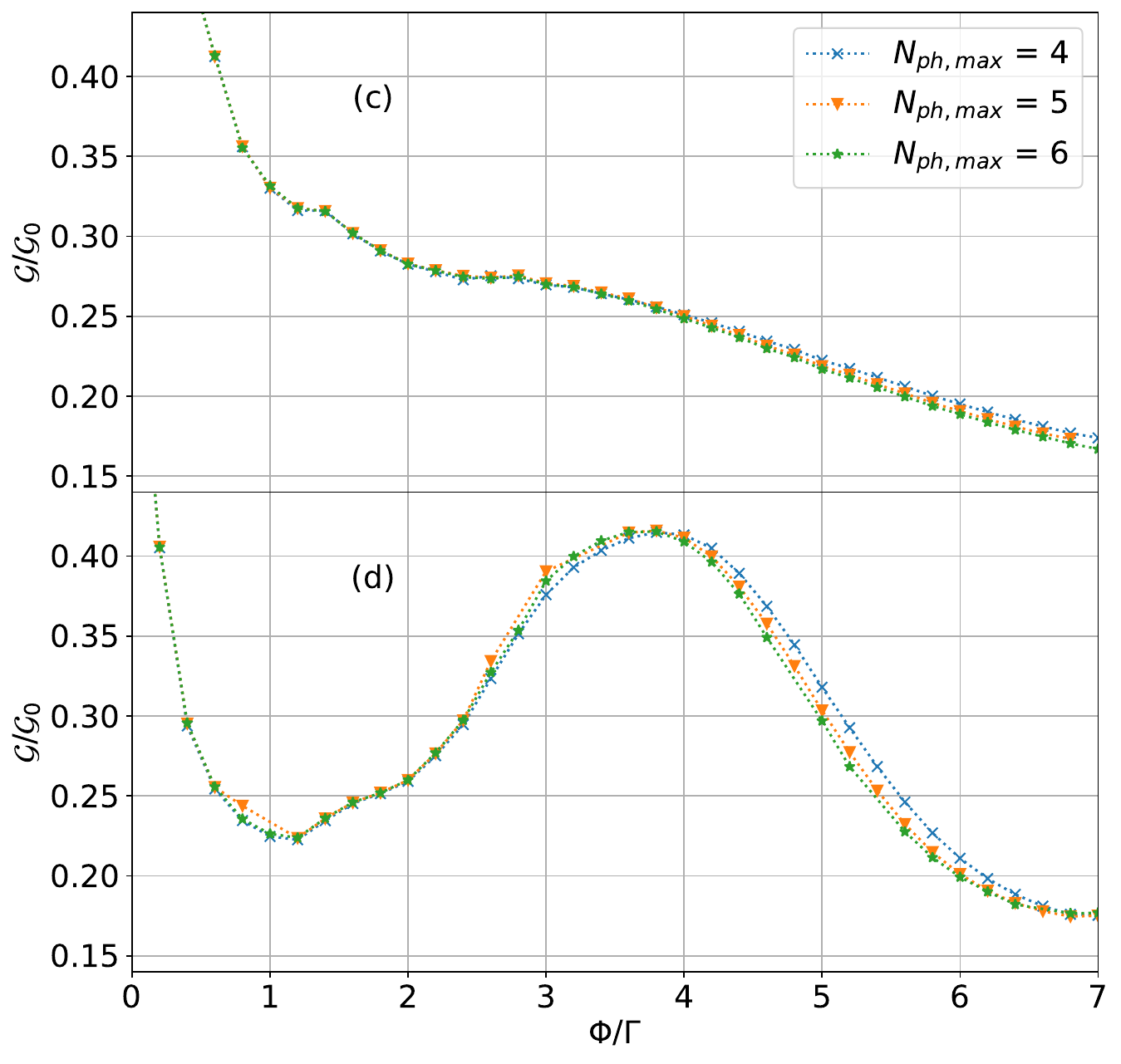}
    \caption{
    Non-equilibrium conductance as a function of voltage evaluated  with different values of the phonon cutoff.
    Results are obtained with the rotation CI (a) and (b) or with the offset CI (c) and (d). (a) and (c) show results with strong lead coupling, $\gamma_\lambda(\omega = 0)/\Gamma = 1.0$, (b) and (d) with weak lead coupling $\gamma_\lambda(\omega = 0)/\Gamma = 0.49$. Remaining parameters are $g/\Gamma = 1.1$, $\omega_\text{b} = 1.5$, $U/\Gamma = 6$, $\varepsilon_\text{imp} = -U/2 + 2g^2/\omega_\text{b}$, $T/\Gamma = 0.05$, $N_\text{ph,max} = 5$. The rotation CI results do not converge for voltages close to zero and to the phonon frequency.}
    \label{fig:non_eq_cond_t_sweep}
\end{figure}

Having shown that in equilibrium the offset CI performs better than the rotation CI, especially at low energies, we now study if this is the case also in non-equilibrium. We consider the differential conductance as a function of applied voltage, and start by investigating the convergence behavior as a function of the number of included phonons. We considered several values of the system-environment coupling ($t'_\lambda$) which we describe in terms of $\gamma_\lambda(\omega = 0) \propto t'^2_\lambda$ (see Eq.~\eqref{eqn:conductance}) for easier notation. Decreasing $\gamma_\lambda(\omega = 0)$ is observed to give rise to a phononic feature at the phonon frequency.

Both transformations converge rapidly for stronger system-environment coupling. In the case of weaker coupling, which is when the phononic feature appears, the rotation CI fails at low voltages, and does not converge at the phonon frequency. The offset CI on the other hand converges almost immediately over the full voltage range. Therefore, especially in non-equilibrium, the rotation does not prove to be advantageous. We expect that a (possibly partial) Lang-Firsov transformation~\cite{ko.lo.11, lo.ko.09, ho.li.07} would bring improvements; this will be implemented in future work.

\begin{figure}[t] 
\centering
\includegraphics[scale=0.35]{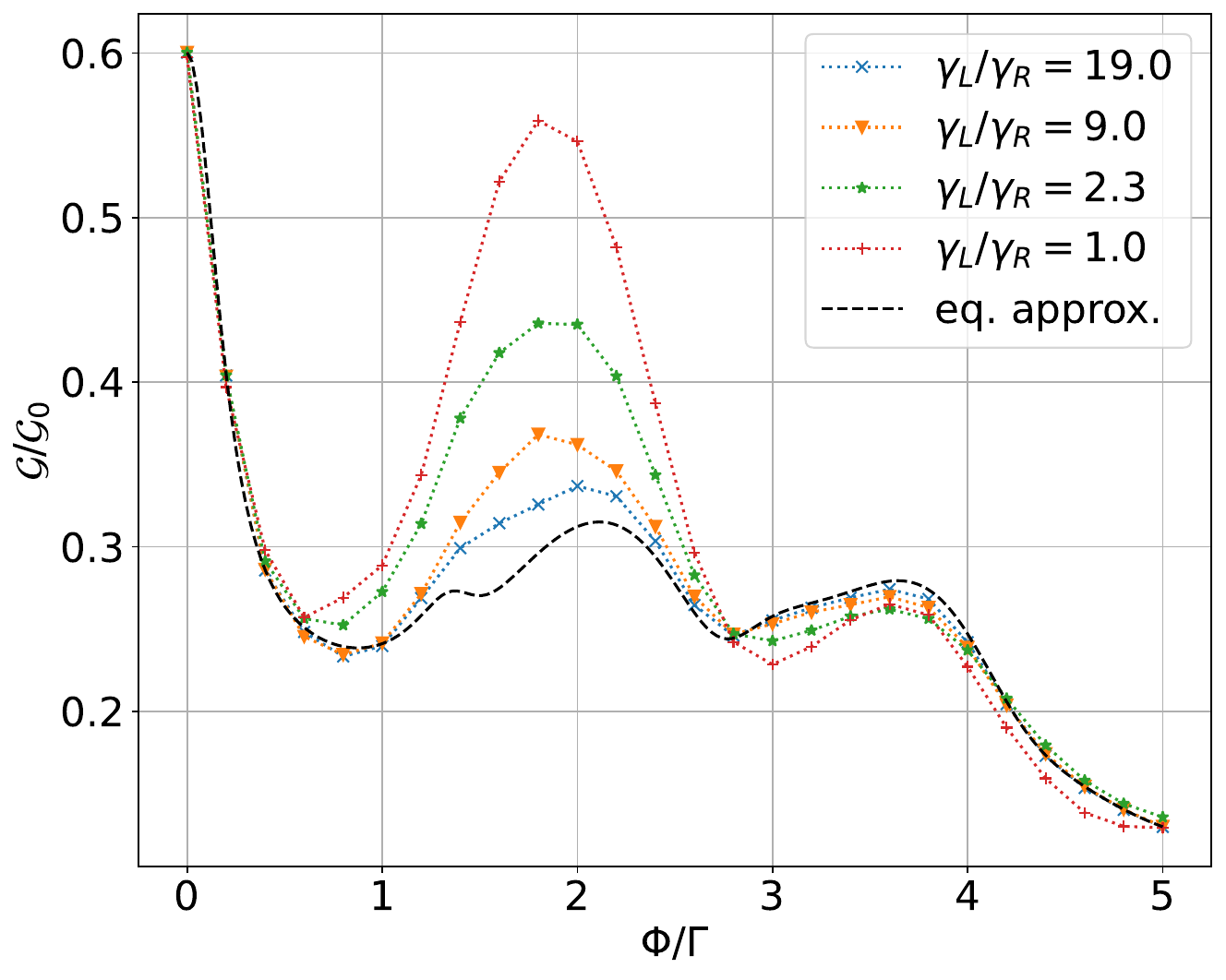}
\caption{Conductance variation with voltage for diverse $\gamma_\text{L}/\gamma_\text{R}$ ratios. Emphasis on asymmetry effects: in pronounced asymmetric scenarios, the differential conductance aligns closely with the equilibrium approximation, but as symmetry increases, non-equilibrium influences become evident. Key parameters: $\mu_\text{L} = 0$, $\mu_\text{R} = \phi$, and $\gamma_R(\omega = 0)/\Gamma + \gamma_L(\omega = 0)/\Gamma = 1.28$. All other parameters are consistent with those in Fig.~\ref{fig:non_eq_cond_t_sweep}}
\label{fig:non_eq_cond_sym_asym}
\end{figure}

We also investigated the transition from very asymmetric lead coupling ($\frac{\gamma_\text{L}}{\gamma_\text{R}} \gg 1$) to the symmetric case. We use the offset CI here, since it proved to be superior to the rotation CI. For $\frac{\gamma_\text{L}}{\gamma_\text{R}} \gg 1$, the finite-bias conductance can be obtained using $G^\text{R}$ from a zero-bias calculation\footnote{This can be seen in the following way: In general, changing the voltage leads to a different Keldysh component of the hybridization function, which then alters $G^\text{R}$ at the impurity. However, provided  the lead in which the chemical potential is changed is weakly coupled, $\Delta^K$ does not change with $\phi$, and consequently neither does $G^\text{R}$. Then the second term in Eq.~\eqref{eqn:conductance} becomes zero, and the first only contains $G_R(\omega, \phi = 0)$.}. Figure \ref{fig:non_eq_cond_sym_asym} shows the nonlinear conductance for a range of $\frac{\gamma_\text{L}}{\gamma_\text{R}}$ using the offset CI. The finite bias is applied by choosing  $\mu_\text{L} = 0$ and $\mu_\text{R} = \phi$. For large asymmetry $\frac{\gamma_\text{L}}{\gamma_\text{R}} = 19$ we see that the conductance is very close to the one obtained by the equilibrium approximation with $\frac{\gamma_\text{L}}{\gamma_\text{R}} \rightarrow \infty$. As we approach equal coupling of the leads, the non-equilibrium effects become apparent and the shape of the conductance curve changes significantly. The vibrational features are expected at $\Phi \sim \omega_b$, while the Coulomb peak is expected at $\Phi \sim U/2 = 3\Gamma$ \cite{mitra2004,paaske2005,rourabas2013}.

\begin{figure}[t] 
\centering
\includegraphics[scale=0.35]{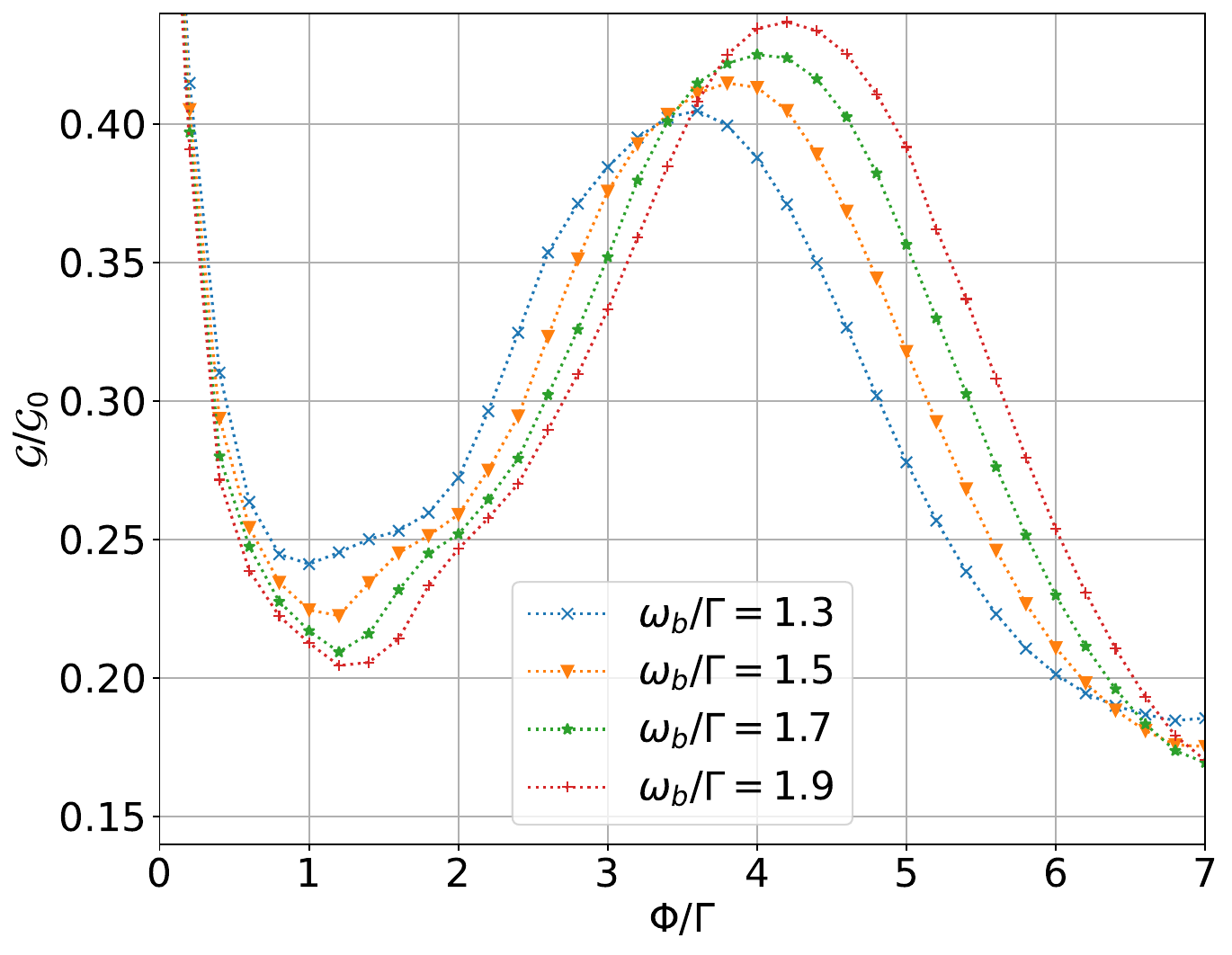}
\caption{Variation in conductance with voltage for various $\omega_b$ values. As the phonon frequency is increased the phononic feature moves to the right, which is to be expected. The Hubbard peak also gets shifted, since the attraction felt by the electrons from the phonons decreases with increasing $\omega_b$. The coupling to the environment is $\gamma_\lambda(\omega = 0)/\Gamma = 0.49$ and all other parameters are in line with Fig.~\ref{fig:non_eq_cond_t_sweep}.}
\label{fig:sweep_wb}
\end{figure}

Finally, we investigate the effect the phonon frequency has on the phononic feature in the fermionic spectrum, as well as on the Hubbard peak. Again we use the offset CI. Figure \ref{fig:sweep_wb} shows that the feature moves with $\omega_\text{b}$, as one would expect, and that it is slowly "absorbed" in the Hubbard peak. The Hubbard peak itself is shifted to higher bias, since increasing $\omega_\text{b}$ decreases the phononic attraction felt by the electrons.

In conclusion, we showed that the method discussed in Ref.~\cite{dz.ko.14} doesn't seem to provide any improvement of the accuracy of the CI calculation with respect to a simple shift in the phononic operators. This is the case especially in non-equilibrium, but also in equilibrium where the phonons have a well-defined temperature. Nevertheless, the equilibrium results presented coincide well with NRG, and the offset CI converges well in non-equilibrium as the phononic Hilbert space is increased. 

\section{\label{sec:conclusion}Conclusion}

We have presented a solver for the Anderson-Holstein problem out of equilibrium. After mapping the Hamiltonian to an auxiliary impurity model in the form of a Lindblad equation and passing to the superfermion representation, we tackled the resulting problem using configuration interaction approach. In the electron sector we use a basis extending up to three particle-hole excitations, while in the phonon sector we attempted to optimize the basis using a shift and rotation, as introduced in Ref~\cite{dz.ko.14}. The method was benchmarked against the numerical renormalization group results in equilibrium, then tested out of equilibrium. We find that in equilibrium the shift as well as the additional rotation allow for high accuracy, using only a small phononic Hilbert space. However, the rotation as discussed in Ref.~\cite{dz.ko.14}, any advantage beyond purely shifting the operators, and in some cases it even introduces instabilities. Therefore, taking into account the effects of the fluctuationg fermionic density on the phonon field
doesn't seem to provide any advantage in the parameter region we have been considering. This becomes even more apparent in non-equilibrium, where the rotation leads to convergence problems in the number of phonons for some parameter regimes. Using the shifted operators, on the other hand, gives quickly converging results. We also use the offset CI to show the propagation of the phononic feature, as well as the Hubbard peak, due to changing phonon frequency. Our solver is thus capable of tackling the problem in the very demanding regime of large bias and strong interactions using only a shift in the phononic operators. Further improvements would consist of implementing the Lang-Firsov transformation. This could potentially lead to a capable solver for interpreting experimental bias spectra.

\begin{acknowledgments}
This research was funded by the Austrian Science Fund (Grant No. P 33165-N) and by NaWi Graz. The results have been obtained using the Vienna Scientific Cluster and the L/A-Cluster Graz. To compute the many-body Lindblad matrix we use the QuSpin library~\cite{wein.23}
\end{acknowledgments}

\bibliography{references_database, additional_citations}

\end{document}